# Segregation of the *Eu* impurity as function of its concentration in the melt for growing of the lead telluride doped crystals by the Bridgman method

D.M. Zayachuk[a,*], O.S. Ilyina[a], A.V. Pashuk[a], V.I. Mikityuk[b], V.V. Shlemkevych[b], A. Scik[c], and D. Kaczorowski[d]

[a]Lviv Polytechnic National University, 12 Bandera St, 79013 Lviv, Ukraine
[b]Yuri Fedkovich Chernivtsy National University, 2 Kotsyubynskoho St, 58012 Chernivtsy, Ukraine
[c]Institute of Nuclear Research, Hungarian Academy of Sciences, H-4026 Debrecen, Bem ter 18/C Hungary
[d]Institute of Low Temperature and Structure Research, Polish Academy of Sciences, 50-950 Wroclaw 2, P. O. Box 1410, Poland

**Abstract**

Behavior of a rare earth impurity of *Eu* in the *PbTe* single crystals grown by Bridgman method from the melt with different initial concentration of impurity $N_{Eu}^{int}(ml)$ about $1 \cdot 10^{20}$ cm$^{-3}$ and less is investigated with X-ray fluorescent element analysis, secondary neutral mass spectroscopy (SNMS), and magnetic measurements. The impurity distributions along and across of the doped ingots are established. It is revealed that doping impurity enters into the bulk of doped crystals only if its initial concentration in the melt is high enough, approximately $1 \cdot 10^{20}$ cm$^{-3}$. If this concentration is lower, about $1 \cdot 10^{19}$ cm$^{-3}$ and less, the doping *Eu* impurity is pushed out onto the surface of doped ingot. The thickness of the doped surface layer is estimated to be in the order of several microns or somewhat more. The longitudinal distribution of *Eu* impurity along the axis of doped ingot – for $N_{Eu}^{int}(ml) = 1 \cdot 10^{20}$ cm$^{-3}$, as well as the transverse one in the surface layer where entire doping impurity is pushed out – for $N_{Eu}^{int}(ml) = 1 \cdot 10^{19}$ cm$^{-3}$, are strongly non-monotonic. Possible reasons for this unusual behavior of *Eu* doping impurity during the growth of *PbTe*:*Eu* crystals from the melt are analyzed.

## 1. Introduction

Doping of semiconductor crystals by foreign impurities has always been the main technological technique for the task-oriented control of their electrophysical parameters such as type and magnitude of conductivity, concentration of free charge carriers, their mobility etc. Many intrinsic and extrinsic factors such as state diagram, temperature gradient at the front of crystallization, crystal growth rate, presence of external physical fields, among others, affect the behavior of doping impurity at the transition from liquid to solid phase during crystallization of ingot and thus determine the outcome and effectiveness of doping. Therefore the processes of impurity segregation in crystals growing from the doped melts are constantly in the focus of research [1-6].

Among the impurities that have been used for a long time for controlling of physical properties of the semiconductor crystals and films, specifically of the IV-VI one, are the impurities of rare earth elements (REE). Numerous studies have shown that doping of the IV-VI crystals with the rare earth impurities meets a lot of problems. In practice it is very difficult to grow the doped crystals containing only the single impurity centers. Usually they contain also the pairs and triplets of the impurities as well as more complicated formations [7-9], and the complexes of the REE with Oxygen [10, 11]. Moreover the rare earth impurities are strongly non-uniformly distributed along the doped crystals during their growth from melt [12-14]. In this article we present the results of investigation of the *Eu* impurity segregation during growth of the lead telluride doped crystals from melt by the Bridgman method as function of initial Europium concentration in the melt.

## 2. Crystal growth and experiment

The *PbTe: Eu* crystals were grown by Bridgman method from the melts of the high-purity initial components. Europium impurity has been introduced into the crystals during the growth. Three different initial concentration of *Eu* were used: $1 \cdot 10^{20}$, $1 \cdot 10^{19}$ and less than $5 \cdot 10^{18}$ cm$^{-3}$. The crystal ingots have a conical-cylindrical shape. The ingot length was 30 mm, the diameter of their cylindrical part 10 mm, the ratio between lengths of the cylindrical and conical parts of ingots about 2:1.

Finding the longitudinal and transverse distributions of doping impurity of Europium in the grown crystals was the main goal of this investigation. Both the bulk and surface *Eu* concentration in the doped ingots was determined with X-ray fluorescent element analysis using the Expert 3L analyzer with semiconducting PIN-detector on thermoelectric cooling. The impurity depth distributions in the surface layers were investigated with secondary neutral mass spectrometry (SNMS). (Dear Attila. Please, add here everything you consider necessary relatively the SNMS experiment). The control of the impurity entry into the ingot crystallized from the melt with the lowest initial impurity concentration was performed with magnetic measurements having a high sensitivity to magnetic impurities. Magnetic measurements were performed at the low temperatures in the range of about 1.7 – 10 K and in applied magnetic fields up to 5 T using a Quantum Design MPMS-5 superconducting quantum interference device (SQUID) magnetometer.

## 3. Experimental results

The experimental investigations have shown that behavior of *Eu* doping impurity in the crystals *PbTe:Eu* growing from the doped melts is extremely sensitive to the initial concentration of *Eu* in the melt $N_{Eu}^{\text{int}}(ml)$ and to the crystal growth process conditions. Thus the behavior of impurities consistently manifests some patterns are closely associated with the magnitude of the initial impurity concentration in the melt.

## 3.1. Initial concentration of *Eu* impurity in the melt $10^{20}$ cm$^{-3}$

With this initial impurity concentration in the melt the quite extended area of the doped crystal ingot (about 2/3 length from the beginning) can be obtained. In this case: (i) the impurity concentration in the beginning of the doped ingot is always higher than its initial concentration in the melt; (ii) the distribution of impurity along the growth axis of the ingot is non-monotonous – during the crystallization process the *Eu* concentration in the ingot first increases, reaches the maximum, and then starts decreasing. In the end of the ingot (approximately 30-35 % of its length) impurity concentration is so low that it cannot be detected by implemented quantitative analysis methods; (iii) across the cylindrical ingot part the impurity concentration is practically constant in the bulk of the crystal, but significantly increases in the surface layers. These patterns of *Eu* impurity behavior are presented in Fig. 1.

Fig.1

## 3.2. Initial concentration of *Eu* impurity in the melt $10^{19}$ cm$^{-3}$

The most characteristic features of the *Eu* impurity behavior at this level of impurity concentration are the following: (i) impurity is distributed only on the surface of doped ingot and is absent in its bulk; (ii) impurity is distributed throughout the whole surface of the ingot, from its beginning to the end, (iii) the impurity concentration on the surface of the crystallized ingot significantly (by the order of magnitude or more) exceeds its magnitude in the initial melt; (iv) the impurity distribution in the surface layers is very sensitive to the actual technological conditions of the crystal growth process, and in general can change significantly from ingot to ingot under the same conditions of their growth (Fig. 2).

Fig. 2

In order to estimate the thickness of the doped surface layer the depth distribution of *Eu* concentration inwards from the surface of the doped ingot was measured with SNMS. The results for the example of ingot 2 (from Fig. 2) are shown in Fig. 3.

Fig. 3

It is evident that the *Eu* impurity is distributed in a very thin surface layer of the doped ingot. The layer thickness is about 7-10 microns and decreases towards the end of the ingots. The transverse distribution of the *Eu* impurity shows non-monotonous character more clearly than the longitudinal one (Fig. 1a). The *Eu* concentration rapidly grows inwards from the surface. At a depth of about 100-200 nm it reaches maximum value, which by a factor of 4 to 5 exceeds the *Eu* concentration on the surface. After reaching a maximum the *Eu* concentration rapidly decreases in depth of the crystal and tends to zero.

## 3.3. Initial concentration of Eu impurity in the melt less than $5 \cdot 10^{18}$ cm$^{-3}$

At this initial *Eu* impurity concentration in the melt its content in the crystallized ingot was lower than the sensitivity of the X-ray fluorescent element analysis. Therefore, control of the impurities entry in the ingot was performed by measuring of the magnetization. Based on the patterns obtained for crystals with $N_{Eu}^{int}(ml) = 1 \cdot 10^{19}$ cm$^{-3}$, one would expect that in this case the doping impurity is distributed mainly in the surface layers of the doped ingot. Therefore the powder sample from the ingot surface was prepared for the experiments. As thin as possible surface layers were removed mechanically. Both the field dependence of magnetization (at 1.72 K) and temperature dependence of magnetic susceptibility (in the magnetic field of 300 Oe) of the powder sample are shown in Fig. 4.

Fig. 4

The fact that the sample is in paramagnetic state in a significant range of magnetic fields and its paramagnetism sharply decreases as temperature increases demonstrates that the surface layer of ingot contains Europium.

**4. Discussion**

The first conclusion which can be drawn from the results of performed research is that the coefficient of segregation of *Eu* impurities $K_S(Eu)$ is greater than one. This is quite natural, since the melting point of *EuTe* is higher than that of *PbTe*. However, if $K_S > 1$, the impurity concentration should decrease in the direction of crystallization of the doped ingot while in experiment the non-monotonic distribution of impurity with a pronounced maximum was observed. (Fig.1*a*). It should be especially emphasizes that such nonmonotonicity observed both for longitudinal (Fig. 1a) and transverse (Fig. 3) distributions of impurity concentration.

A sharp trend of the impurity concentration to zero in the area of its decrease towards the ingot end is another specific feature of the longitudinal impurity distribution (Fig. 1a). In the last one third of the doped ingot doping impurity is detected neither in the bulk nor on the surface of the ingot, where impurity is intensely pushed out (Fig.1b). However, if the initial *Eu* concentration in the melt is low and the doping impurity is pushed out onto the ingot surface in the process of its crystallization the impurity is distributed over the entire surface up to the end of the ingot (Fig. 2).

It is well known that distribution of impurities in the doped crystals grown from melts is determined by the phase diagram of the system "crystal matrix − impurity". In case of linear approximation of dependence of liquidus $T_L$ and solidus $T_S$ temperatures on impurity content $x_{imp}$, the segregation coefficient is constant. For nonlinear approximation of $T_L(x_{imp})$ and $T_S(x_{imp})$ dependencies the segregation coefficient becomes dependent on the concentration of impurities in the melt. Naturally, this significantly changes the impurity distribution along the axis of the doped ingot. However, no matter how strongly $T_L$ and $T_S$ are dependent on $x_{imp}$, the distribution will not be non-monotonic. This suggests that the real longitudinal and transverse distributions of *Eu* impurity in the *PbTe:Eu* crystals are the result of superposition of at least two different physical processes.

One of these processes is, naturally, the segregation of *Eu* impurity. Since $K_S(Eu) > 1$, the contribution of this process to the final outcome is dominant in the decreasing region of concentration profiles of the impurity distribution. Under this premise, we have analyzed the decreasing region of the doping impurity distribution of the sample in Fig. 1. Its multinomial extrapolation to the beginning of ingot provides the magnitude of the impurity concentration of about $(2.5 \div 3.0) \cdot 10^{20}$ cm$^{-3}$, depending on the multinomial degree. Averaging the extrapolation data, the segregation coefficient magnitude $K_S(N_{Eu}(ml) = 1 \cdot 10^{20}$ cm$^{-3}) = 2.75$ was chosen as a starting point for the analysis. Comparison of experimental and calculated distributions for such $K_S$ magnitude is shown in Fig. 5. One can see that the segregation coefficient independent of concentration cannot explain the specificity of the experimentally observed behavior of impurities in the solid phase, in particular the sharp decrease of the impurity concentration in the second half of doped ingot and its absence in the end of the ingot. Therefore, the same data were analyzed with quadratic approximation of the state diagram:

$$T_S(x_S) = T_0 + ax_S + bx_S^2$$
$$T_L(x_L) = T_0 + cx_L + dx_L^2, \quad (1)$$

where $x_L$ and $x_S$ are the impurity contents in the liquid and solid phases, respectively. In this approximation:

$$K_S = \frac{a}{2bx_L}\left(\sqrt{1 + \frac{4b}{a}\left(\frac{c}{a}x_L + \frac{d}{a}x_L^2\right)} - 1\right). \quad (2)$$

As for a given $x_L$ the magnitude of $K_S$ is determined by three parameters $b/a$, $c/a$, i $d/a$, and we have only one starting point for the analysis – the initial concentration of impurities in the melt $N_{Eu}^{int}(ml) = 1 \cdot 10^{20}$ cm$^{-3}$ ($x_L = 0.00337$), then further analysis was carried out for the two limiting cases where the nonlinear on the impurity concentration is only one of the two lines of the phase diagram – either the liquidus line (then $b/a = 0$) or the solidus one (then $d/a = 0$). The ratio between other parameters ($c/a$ and $d/a$ for the first case; $c/a$ and $b/a$ for the second one), which determines the magnitude of $K_S$, was chosen so as to meet a condition $K_S(x_L = 0.00337) = 2.75$.

Fig.5.

One of the model impurity distributions along the doped ingot for the case of nonlinear solidus line of the "compound – impurity" phase diagram is shown in Fig. 5. As can be seen, it is very different from the distribution for the case of the independent of the impurity concentration segregation coefficient and accurately describes the rapid decrease of concentration in the second half of doped ingot and absence of impurity at its end. Thus, there is a solid ground to claim that the obtained coordinate distribution of $Eu$ impurity in the doped ingot for the case of $N_{Eu}^{int}(ml) = 1 \cdot 10^{20}$ cm$^{-3}$ definitely points out that the $Eu$ segregation coefficient of in $PbTe$ depends on the impurity concentration in the melt. For the model distribution presented in Fig. 5 with the parameter magnitudes of the state diagram (1) $c/a = 79.2$, $d/a = 0$ and $b/a = 3000$, this dependence is shown in Fig. 6.

Fig.6.

In our opinion, another one of the two processes determining the resulted distribution of the *Eu* impurities in the *PbTe:Eu* crystals mentioned above is formation of the complexes of doping and uncontrolled background impurities, predominantly Oxygen, during crystallization. This is suggested by the following well-known facts. Firstly, REE impurities are chemically very active and have a strong gettering action (see eg. [15]). Secondly, during growth of the *PbTe* crystals from the melt doped with REE, the "REE impurity – Oxygen" complexes are formed with high probability at the beginning of doped crystal and absent at its end [11].

A formation mechanism of the "impurity *REE* – Oxygen" complexes in a *PbTe* doped crystal matrix is unknown. It can be suggested that they are formed at the solid – liquid interface in front of the front of crystallization, where the temperature of melt is the lowest, and thus conditions for the complex formation in melt is the most favorable. The probability of the complex formation is proportional to both the concentration of Europium and the concentration of Oxygen in the melt at the front of crystallization. The complexes, especially the large ones, formed in front of the front of crystallization will be poorly integrated into the crystallizing matrix, and pushed back from it towards the liquid phase. Thus, on the front of crystallization these complexes will behave like impurities with segregation coefficient less than one.

Taking this into account we were modeled the concentration profiles of *Eu* doping impurity in the *PbTe:Eu* ingot as the results of superposition of two different mechanisms for its entry into the crystallizing matrix. These mechanisms are described by two different segregation coefficients – $K_S(Eu) > 1$, and $K_S(comp) < 1$. Also it is taken into account that the melt is cleared from Oxygen if Europium getters it and forms the complexes in front of the front of crystallization. First, clearing of the melt will promote the decay of the large complexes formed before and pushed back to the liquid phase during crystallization. Second, clearing of the melt reduces the probability of formation of the new complexes during crystallization of remaining melt. As a result, during crystallization of doped ingot the contribution of atomic Europium in the total impurity concentration in the crystallizing layer increases, and the contribution of Europium as a constituent of complexes decreases.

For simulation of possible impurity profiles as a result of superposition of mentioned two processes let's denote a portion of atomic *Eu* in the melt at the front of crystallization as *m*. The magnitude of *m* will constantly increase in the process of crystallization and tend to unity towards the ingot end. The rate of change of *m* during the crystallization process will depend on both the initial concentration of doping impurities in the melt and the initial concentration of uncontrolled Oxygen. Taking this into consideration, dependence *m(x/L)* was approximated by the following relation for model calculations:

$$m(x/L) = 1 + A(1 - \exp(\frac{1 - x/L}{B})), \quad (3)$$

where *x* is a coordinate along the axis of ingot growth, which changes from 0 to the length of ingot *L*. The model dependences of *m(x/L)* for different magnitudes of parameter *B* under the assumption that all impurity at the front of crystallization exists in the form of complexes at the beginning of crystallization process, and in the atomic form at the end of this process is shown in Fig. 7.

Fig.7.

Different magnitudes of parameter *B* in relation (3) corresponds to different contents of uncontrolled Oxygen in the melt. When the later increases, the parameter *B* increases too. Naturally the faster magnitude of m increases and reaches unity during the crystallization of the ingot, the lower is the content of uncontrolled oxygen in the melt.

Using the assumption about two different processes of *Eu* segregation in *PbTe:Eu* crystals we have tried to simulate the two most fundamental features of the longitudinal concentration profile of impurities, namely the maximum of concentration and its coordinate position in the ingot. Concentration dependence of $K_S(Eu)$ was chosen the same as in Fig. 6.

Predetermined concentration of *Eu* at the beginning of doped ingot can be obtained by combining of magnitude of both parameters *m(x/L = 0)* and $K_S(comp)$. Then the resulting impurity profile depends on the magnitude of parameter *B*. Examples of this dependence for two different combinations of *m(x/L=0)* and $K_S(comp)$ are shown in Fig. 8.

Fig.8.

As one can see, for both combinations of parameters *m(x/L=0)* and $K_S(comp)$, giving the same magnitude of the impurity concentration at the ingot beginning, the impurity distributions in the solid phase are very close, if parameter *B* (background Oxygen impurity content in the melt) is the same. At the same time, if parameter *B* is lower, the concentration profile has a more pronounced maximum, which is located closer to beginning of the doped ingot.

Comparison of the experimental *Eu* concentration profile with the model dependences shown in Fig. 9 for the case of B = 0.2 for which the experimental data could be best reproduced by model curves in terms of the above problem. This reproduction is good enough and naturally can be achieved with different sets of parameters $K_S(comp)$ and *m*. In our opinion, this result may be considered as evidence supporting the abovementioned hypothesis concerning the mechanism of doping of *PbTe* with *Eu* impurity during growth of ingots from doped melts.

Fig.9.

The nature of nonmonotonicity of the radial distributions of impurities in the crystals with a low initial concentration of Europium in the melt (Fig.3) is of course the same as the nature of nonmonotonicity of the longitudinal distribution of impurities in the crystals with a high initial concentration of Europium (Fig. 1a). Unfortunately, it is impossible to simulate them in a simple way like longitudinal distribution. Firstly, the radial redistribution of *Eu* in the surface layers occurs under strongly non-equilibrium conditions due to rapid heat removal through lateral surfaces. Secondly, this redistribution occurs in two mutually perpendicular temperature gradients – the axial, created by the furnace for crystal growth, and the radial created by heat rejection through the lateral surfaces. Therefore, there are too many free parameters for the simulation. Nevertheless, some qualitative conclusions could be drawn. Comparing the longitudinal (Fig.1a) and the transverse (Fig.3) distributions with the model ones (Fig.8) it can be suggested that the transverse distribution corresponds to the much lower magnitude of parameter *B* than the longitudinal one. This means that Oxygen content in the melt with the initial concentration of *Eu* $N_{Eu}^{int}(m) = 1 \cdot 10^{19}$ cm$^{-3}$ (Fig.3) is lower than when $N_{Eu}^{int}(m) = 1 \cdot 10^{20}$ cm$^{-3}$ (Fig.1a). This suggests that majority of uncontrolled Oxygen enters the melt together with the *Eu* impurity during the doping of crystals. Such result is expected because obtaining the pure rare earth elements is extremely complicated technical task.

Magnetization and magnetic susceptibility data shown in Fig. 4 allow estimating the Europium concentration in the surface layer of ingot grown from the melt with $N_{Eu}^{int}(m) < 5 \cdot 10^{18}$ cm$^{-3}$. A coincidence of magnetization of the sample under up-down change of magnetic field indicates that there are no inclusions of the ferromagnetic *EuO* phase in the investigated sample. Recently we have shown [9] that the main contribution to the magnetization of the crystal

*PbTe:Eu* are done by the crystal matrix, the single *Eu* centers and its NN (nearest neighbor) and NNN (next nearest neighbors) pairs within the Europium and Oxygen complexes. Based on this and using magnitudes of the exchange integrals $J_1/k_B = +0.056$ K and $J_2/k_B = -0.13$ K [9] for ferromagnetic and antiferromagnetic interaction between NN and NNN pairs respectively, we were performed analytical treatment of magnetization and magnetic susceptibility data. The standard relations given in [9] were used for calculations. The results of treatment are shown in Fig. 10.

Fig.10.

The best coincidence between experimental and calculated data for both magnetization (Fig. 10a) and magnetic susceptibility (Fig. 10b) was achieved under the assumption that all *Eu* in the surface layers forms simple NN and NNN complexes in approximately equal proportions of its total concentration in the sample about $2.35 \cdot 10^{18}$ cm$^{-3}$. Obviously, this magnitude is much lower than the real averaged *Eu* concentration in the surface layer, where *Eu* is pushed out during the doped ingot growth, as the thickness of the surface layer, which was mechanically removed for the preparation of powder sample was at least some tens of microns, and according to the data in Fig. 3 for the low $N_{Eu}^{int}$(*ml*) *Eu* impurity is distributed in a thin surface layer with a thickness of just a few micrometers.

Another important result of these measurements of magnetic properties is a hint for understanding why the *Eu* impurity is distributed along lateral surface of the doped ingots grown from the melt with high and low initial concentration of doping impurity in the different ways. As shown above the surface layers of the ingot grown from melt with low initial impurity concentration practically does not contain the single *Eu* centers whereas only the single Eu$^{2+}$ centers are identified in the end part of *PbTe:Eu*(*Gd*) ingots grown from the melt with high initial impurity concentration [9, 11]. The former means that in the surface layers where the doping impurity is pushed out during growth of doped crystals from the melt with the low initial impurity concentration there always are suitable conditions for complex formation. The complexes are constantly being pushed out of solid into liquid phase, maintaining the Europium content in the liquid phase, and thus the impurity spreads over the surface up to the end of the doped ingot. Alternatively, if only the *Eu* single centers exist in the end parts of the doped *PbTe:Eu*(*Gd*) ingots grown from the melt with high initial impurity concentration then only the mechanism of atomic *Eu* segregation works and all impurity pull out quickly from the liquid phase since $K_S(Eu)$ is high and increases if impurity concentration in the melt decreases.

**5. Conclusions**

    *Eu* impurity segregation in the lead telluride doped crystals grown by the Bridgman method from melts with different initial concentration of impurity $N_{Eu}^{\text{int}}(m)$ is investigated. X-ray fluorescent element analysis, secondary neutral mass spectroscopy, and magnetic measurements were used for this purpose. It is revealed that distribution of doping impurity in the doped crystal drastically depends on the initial concentration of impurity in the melt. If $N_{Eu}^{\text{int}}(m)$ is about $1 \cdot 10^{20}$ cm$^{-3}$ Europium is distributed over the entire cross section of the doped ingot, and is concentrated in the initial part of the ingot of about 2/3 of its length. If $N_{Eu}^{\text{int}}(m)$ is about $1 \cdot 10^{19}$ cm$^{-3}$ and less the doping *Eu* impurity is pushed out onto the surface of doped ingot and is distributed along the lateral surface of the entire length of ingot. In this case, both the longitudinal (for $N_{Eu}^{\text{int}}(m) = 1 \cdot 10^{20}$ cm$^{-3}$) and transverse (for $N_{Eu}^{\text{int}}(m) = 1 \cdot 10^{19}$ cm$^{-3}$) distributions of impurities are strongly non-monotonic.

    We suggest that non-monotonic distributions of the doping *Eu* impurity are caused by superposition of the two different mechanisms of its entering into crystal from the doped melt. One of them is entering of the single *Eu* atoms with a segregation coefficient more than unity, which strongly depends on the impurity concentration in the melt and increases when this concentration decreases. Another one is entering of *Eu* as a constituent of complexes with Oxygen, which is formed at the solid − liquid interface in front of the front of crystallization. When entering the solid phase, these complexes behave as impurity with a segregation coefficient less than unity.

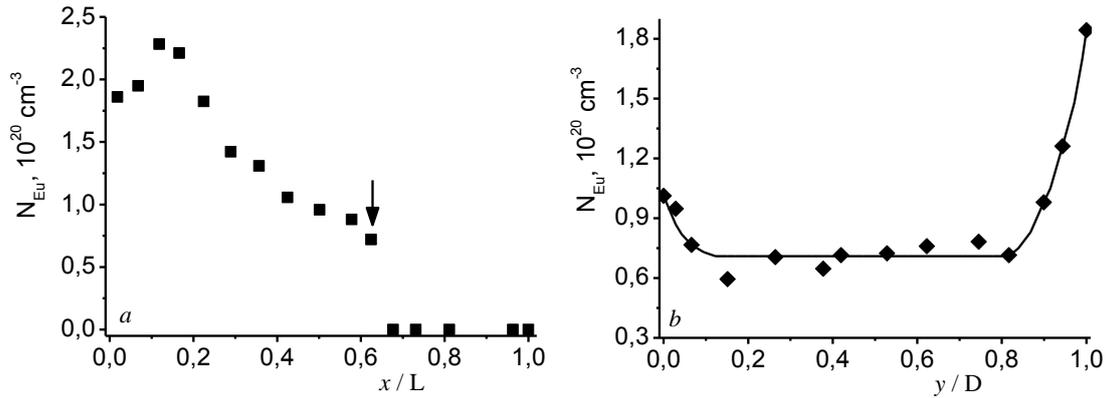

Fig.1. *Eu* doping impurity distributions in the ingot *PbTe: Eu* grown from the melt doped with initial impurity concentration of $1\cdot 10^{20}$ cm$^{-3}$: *a*) along the axis of growth: *b*) across ingot (cross-section indicated by the arrow in (a)). $x$ – longitudinal coordinate, $y$ - transverse coordinate, $L$ – total length of the ingot, $D$ - diameter of the ingot cylindrical part

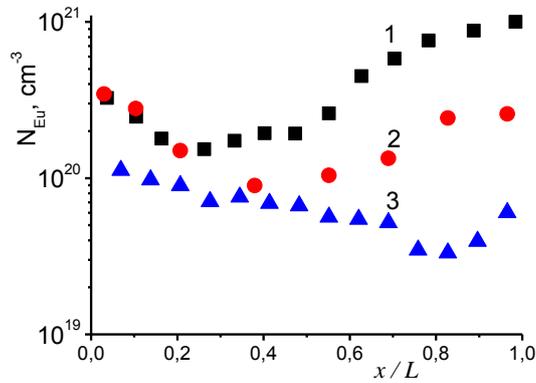

Fig. 2 Longitudinal distribution of the *Eu* doping impurity along the lateral surface of the different *PbTe: Eu* ingots grown from the melt doped with the same initial *Eu* concentration of $1\cdot 10^{19}$ cm$^{-3}$

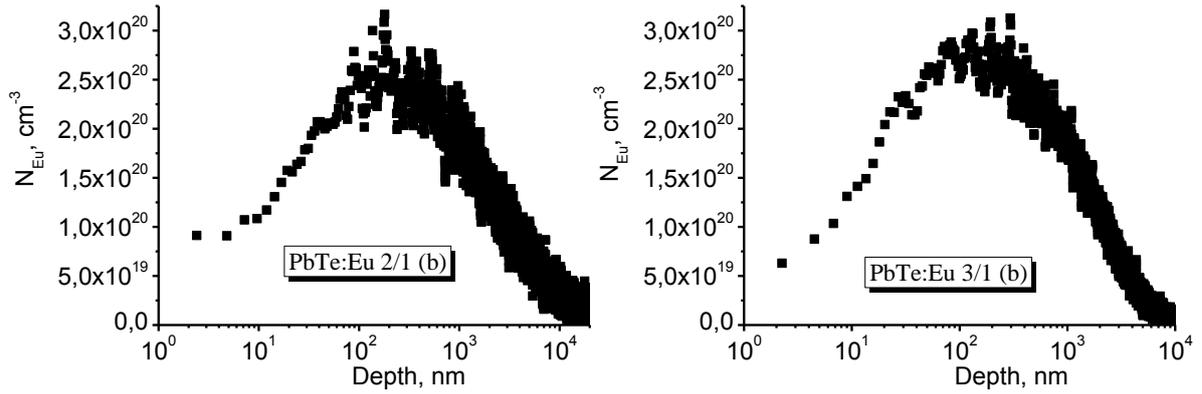

Fig. 3 Depth distribution of *Eu* inward from the surface of the doped ingot with $N_{Eu}^{int}(ml) = 10^{19}$ cm$^{-3}$ (The ingot 2 in Fig.2) at a distance of about 17 (*a*) and 26 mm (*b*) from beginning of the ingot

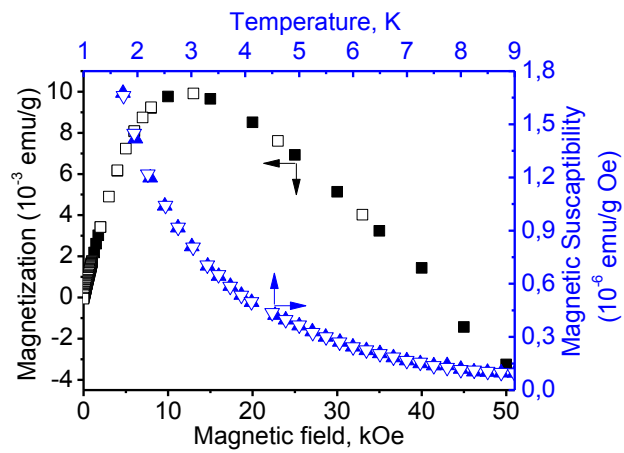

Fig. 4 Magnetization of the powder samples vs. magnetic field at the temperature 1.72 K (black symbols) and magnetic susceptibility vs. temperature at the magnetic field 300 Oe (blue symbols). The solid symbols are the data when the argument (magnetic field or temperature) increases; the open symbols are the data when the argument decreases.

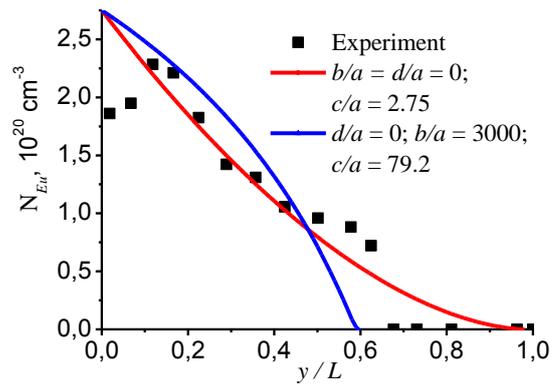

Fig.5. Experimental and simulated distributions of the *Eu* concentration along the ingot grown from the melt with initial impurity concentration $N_{Eu}^{int}(ml) = 1 \cdot 10^{20}$ cm$^{-3}$ for the "compound - impurity" state diagrams with different magnitudes of *a, b, c,* and *d* parameters

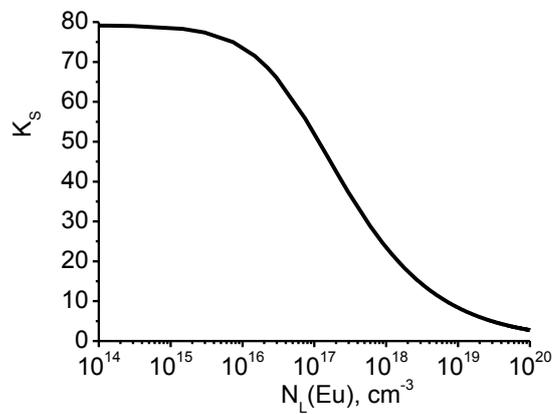

Fig.6. Concentration dependence of the coefficient of impurity segregation for the parameter magnitudes of the phase diagram (1): $c/a = 79.2$, $d/a = 0$, and $b/a = 3000$

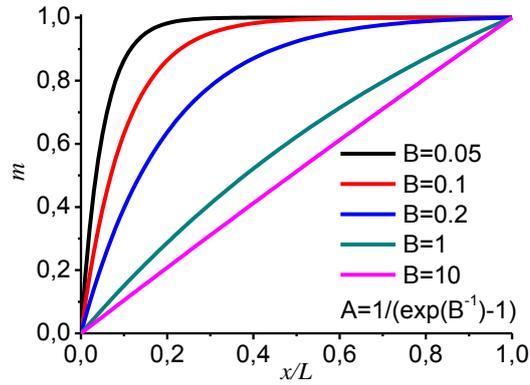

Fig.7. Model curves of parameter *m* given by the relation (3) for different magnitudes of parameter *B* in suggestion that *m(x=0)* = 0, and *m(x=L)* = 1

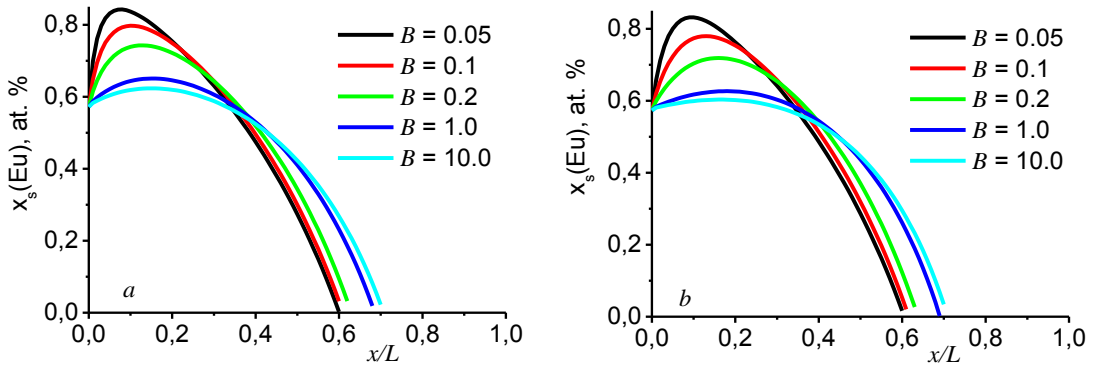

Fig.8. Concentration profiles of doping impurity as a result of superposition of processes of segregation of *Eu* in both the atomic form and the constituent of complexes with Oxygen for different magnitudes of parameters *m(x/L=0)*, *B*, and $K_S(comp)$: *a) m(x/L=0)* = 11.5 %, $K_S(comp)$ = 0.9; *b) m(x/L=0)* = 36.2 %, $K_S(comp)$ = 0.1. Magnitudes of parameter *B* are indicated in the figures.

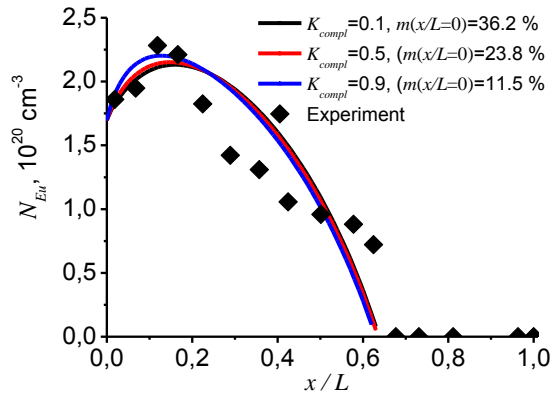

Fig.9. Comparison of the experimental impurity distribution along the axis of *Eu*-doped ingot grown from the melt with initial impurity concentration $N_{Eu}^{int}(ml) = 1 \cdot 10^{20}$ cm$^{-3}$ and model ones calculated for different combinations of $m(x/L=0)$ and $K_S(comp)$ if $B = 0.2$

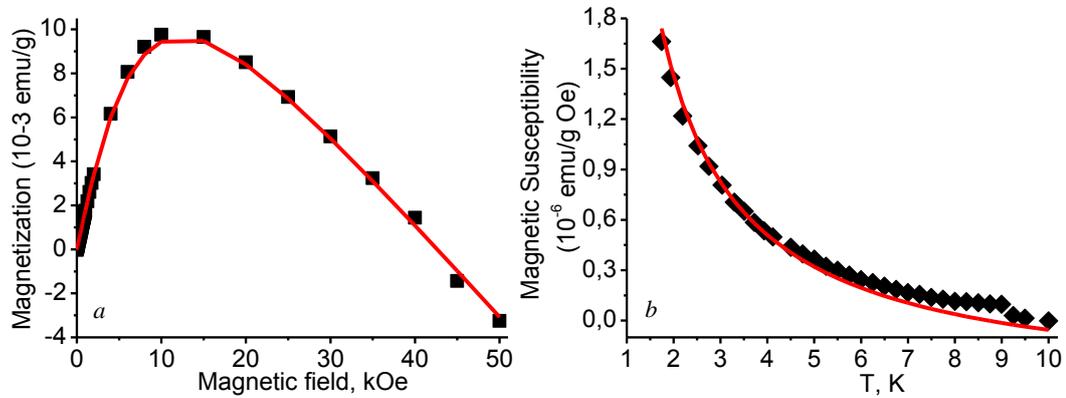

Fig.10. Comparison of experimental data presented in Fig. 4 (symbols) and calculated ones (solid lines). Parameters for calculation: *Eu* concentration in the NN pairs is equal to $1.35 \cdot 10^{18}$ cm$^{-3}$; *Eu* concentration in the NNN pairs is equal to $1 \cdot 10^{18}$ cm$^{-3}$; magnetic susceptibility of the crystal matrix $\chi_{Matrix} = -0.43 \cdot 10^{-6}$ emu/(g·Oe)